\documentclass[12pt]{article}
\usepackage{times}
\usepackage{geometry}
\usepackage{graphicx}
\usepackage[utf8]{inputenc}
\usepackage{enumitem,amssymb}
\usepackage{ragged2e}
\newlist{thematic}{itemize}{8}
\setlist[thematic]{label=$\square$}
\usepackage{pifont}
\newcommand{\cmark}{\ding{51}}%
\newcommand{\done}{\rlap{$\square$}{\raisebox{2pt}{\large\hspace{1pt}\cmark}}%
\hspace{-2.5pt}}

\usepackage{natbib}

\begin{document}
\raggedright
\huge
Astro2020 Science White Paper \linebreak

Planet formation – The case for large efforts on the computational side
 \linebreak
\normalsize

\noindent \textbf{Thematic Areas:} %\hspace*{60pt} \done Planetary Systems \hspace*{10pt} \done Star and Planet Formation \hspace*{20pt}\linebreak
\hspace*{60pt} \done Star and Planet Formation \hspace*{20pt}\linebreak
%$\square$ Formation and Evolution of Compact Objects \hspace*{31pt} $\square$ Cosmology and Fundamental Physics \linebreak
%  $\square$  Stars and Stellar Evolution \hspace*{1pt} $\square$ Resolved Stellar Populations and their Environments \hspace*{40pt} \linebreak
%  $\square$    Galaxy Evolution   \hspace*{45pt} $\square$             Multi-Messenger Astronomy and Astrophysics \hspace*{65pt} \linebreak
  
\textbf{Principal Author:}

Name: Wladimir Lyra	
 \linebreak						
Institution: California State University, Northridge 
 \linebreak
Email: wlyra@csun.edu
 \linebreak
Phone:  +1 818-677-7464
 \linebreak
 
\textbf{Co-authors:} 
%\begin{justify}
Thomas Haworth (Imperial College London), % t.haworth@imperial.ac.uk 
Bertram Bitsch (MPIA), % bitsch@mpia.de
Simon Casassus (Universidad de Chile), % scasassus@u.uchile.cl
Nicol\'as Cuello (Pontificia Universidad Cat\'olica de Chile), % ncuello@astro.puc.cl 
Thayne Currie (NASA Ames), % thayne.m.currie@nasa.gov  
Andras G\'asp\'ar (University of Arizona), % agaspar@as.arizona.edu
Hannah Jang-Condell (University of Wyoming), %, hjangcon@uwyo.edu
Hubert Klahr (MPIA), % klahr@mpia.de 
Nathan Leigh (AMNH/Universidad de Concepcion), %nleigh@amnh.org  
Giuseppe Lodato (Universit\`a di Milano), % giuseppe.lodato@unimi.it 
Mordecai-Mark Mac Low (AMNH/CCA), % mordecai@amnh.org  
Sarah Maddison (Swinburne University of Technology), %  smaddison@swin.edu.au
George Mamatsashvili (Niels Bohr Institute), % George.mamatsashvili@nbi.ku.dk 
Colin McNally (Queen Mary University of London), %, c.mcnally@qmul.ac.uk
Andrea Isella (Rice University), % isella@rice.edu 
Sebasti\'an P\'erez (Universidad de Santiago de Chile), % sebastian.astrophysics@gmail.com
Luca Ricci (CSUN), % luca.ricci@csun.edu
Debanjan Sengupta (NASA Ames), % debanjan@udel.edu 
Dimitris Stamatellos (University of Central Lancashire), % DStamatellos@uclan.ac.uk
Judit Szul\'agyi (University of Z\"urich), % judit.szulagyi@uzh.ch
Richard Teague (University of Michigan), % rteague@umich.edu 
Neal Turner (JPL), % Neal.J.Turner@jpl.nasa.gov 
Orkan Umurhan (NASA Ames), % Orkan.M.Umurhan@nasa.gov
Jacob White (Konkoly Observatory), % jacob.white@csfk.mta.hu 
Al Wootten (University of Virginia). % awootten@nrao.edu
\linebreak

%\end{justify}

\textbf{Co-signers:} 

%\begin{justify}
Felipe Alarcon (University of Michigan), % falarcon@umich.edu
Daniel Apai (University of Arizona), % apai@arizona.edu
Amelia Bayo (Universidad de Valparaiso), % amelia.bayo@uv.cl
Edwin Bergin (University of Michigan), % ebergin@umich.edu
Daniel Carrera (Penn State University), % danielc@psu.edu 
Ilse Cleeves (University of Virginia),  % lic3f@virginia.edu
Asantha Cooray (UC Irvine), %acooray@uci.edu
Gregor Golabek (University of Bayreuth),  % gregor.golabek@uni-bayreuth.de
Oliver Gressel (Niels Bohr Institute), % gressel@nbi.ku.dk 
Mark Gurwell (Harvard-Smithsonian Center for Astrophysics), % mgurwell@cfa.harvard.edu
Sebastiaan Krijt (University of Arizona), % skrijt@email.arizona.edu
Cassandra Hall (University of Leicester), % cassandra.hall@leicester.ac.uk 
Ruobing Dong (University of Victoria), % rbdong@uvic.ca 
Fujun Du (University of Michigan), % fdu@umich.edu
Ilaria Pascucci (LPL), % pascucci@lpl.arizona.edu
John Ilee (University of Leeds), % j.d.ilee@leeds.ac.uk
Andre Izidoro (Universidade Estadual Paulista), % izidoro.costa@gmail.com 
Jes Jorgensen (Niels Bohr Institute), % jeskj@nbi.ku.dk
Mihkel Kama (University of Cambridge), % mkama@ast.cam.ac.uk
Dimitri Mawet (Caltech) % dimitri.mawet@gmail.com
Jinyoung Serena Kim (University of Arizona), % serena@as.arizona.edu
David Leisawitz (NASA Goddard), %david.t.leisawitz@nasa.gov
Tim Lichtenberg (University of Oxford), % tim.lichtenberg@physics.ox.ac.uk
Nienke van der Marel (NRC Herzberg), % astro@nienkevandermarel.com
Margaret Meixner (STScI), % meixner@stsci.edu 
John Monnier (University of Michigan), % monnier@umich.edu
Giovanni Picogna (Ludwig-Maximilians-Universit\"at M\"unchen), % picogna@usm.lmu.de
Klaus Pontoppidan (STScI), % pontoppi@stsci.edu
Hsien Shang (ASIAA), % shang@asiaa.sinica.edu.tw 
Jake Simon (SwRI), %jbsimon.astro@gmail.com
David Wilner (Harvard-Smithsonian Center for Astrophysics), %dwilner@cfa.harvard.edu
  \linebreak
%\end{justify}

%\textbf{Abstract  (optional):}

\pagebreak

%\textbf{Introduction}

\section{Introduction}

\begin{justify}
Planet formation is simultaneously one of the oldest and one of the
newest concerns of human inquiry. “How did the Earth come to be?” is a
question that almost invariably appears in the cosmogonies of the
ancients. In the late modern period it was understood that planets
must form in disks of gas and dust around young stars, a process that
modern astronomy has finally been able to observe in reasonable
resolution and detail. 

Developments in technical capabilities at optical/near-infrared and sub-millimeter wavelengths in the
2010--2019 decade, in particular the advent of the Atacama Large
Millimeter Array (ALMA) and high-contrast adaptive optics have produced spatially resolved observations
of the structures within protoplanetary disks (see reviews by \citealt{Casassus16} and \citealt{Sicilia-Aguilar+16}). This has revealed a plethora of sub-structure, including
rings and gaps \citep{ALMA+15},
spirals \citep{Muto+12,Garufi+13,Benisty+15,Currie+14,Currie+15}, warps \citep{Casassus+15,Cuello+19}, shadows
\citep{Stolker+16,Kama+16,Cuello+19}, cavities \citep{Andrews+11}, and
dust traps \citep{vanderMarel+13,Casassus+18}. For a recent survey, see \citet{Garufi+18}. 
 
These processes are understood under the framework of disk-planet
interaction, a process studied analytically and modeled numerically
for over 40 years \citep{GoldreichTremaine79,LinPapaloizou93,KleyNelson12}. Long a
theoreticians' game, the wealth of observational data has been
allowing for increasingly stringent tests of the theoretical
models. Although the observed structures qualitatively matched the
general predictions from these models, one of the highlights of the
decade of 2010--2019 in planet formation was the attempt to bring the
hydrodynamical models to the level of {\it quantitative} agreement
with the new detailed observations, a task that has been unexpectedly
challenging. Modeling efforts are crucial to support the
interpretation of direct imaging analyses, not just for potential
detections but also to put meaningful upper limits on mass accretion
rates and other physical quantities in current and future large--scale surveys.

The path towards a complete theory of planet formation 
remains elusive. While some processes have been definitely 
observed (such as rings, gaps, spirals, and vortices), and some 
indirectly observed, such as turbulence \citep{Flaherty+15,Teague+18}
other processes are less likely to be directly observed (such as
planet formation via streaming instability and pebble accretion). 

As even more detail is expected with ground based interferometers, 
extremely large telescopes (ELTs) and the James Webb Space Telescope (JWST)
in the next decade (\citealt{Ricci+18} and white papers by Isella,
Currie,  and Jang-Condell), a burning question, and 
the central point of this white paper, is {\it what efforts on the
  computational side are required in the next decade to advance 
our theoretical understanding, explain the observational data, and guide new observations}? 

\end{justify}

\section{Overview of computational planet formation}   

%\begin{figure}
% \begin{center}
%   \resizebox{.7\textwidth}{!}{\includegraphics{scheme.png}}
%   \caption{{\sf Protoplanetary disk schematic highlighting some of the key disk mechanisms and physics we are required to model to capture them (in parentheses). These physical ingredients are hydrodynamics (HD), magnetohydrodynamics (MHD), radiation hydrodynamics (RHD), radiative transfer (RT), chemistry (CHEM), and dust dynamics (DD). The background image is a subset of Hubble observation of R136, credit: NASA, ESA, and F. Paresce (INAF-IASF, Bologna, Italy), R. O’Connell (University of Virginia, Charllotsville), and the Wide Field Camera 3 Science Oversight Committee.}}
%    \label{fig:scheme}
% \end{center}
%\end{figure}

\begin{justify}

The dynamical state of the protoplanetary disk is the fundamental canvas on which the planet formation narrative is
etched. Understanding the evolution of disks, 
the structures that we are observing within them and the planet formation process presents a formidable challenge. Disks are composed of 
material spanning conditions ranging from cold, dense, and molecular, through to diffuse, hot, and ionized. Densities and temperatures vary by about 10 and 3 orders of magnitude, respectively. The gravitational potential from the parent star, self-gravity of the disk, hydrodynamic torques in the disk, radiation from the parent star or other nearby stars, dust, cosmic rays, 
and non-ideal magnetohydrodynamics (MHD) all play important roles \citep[see
reviews by][]{DullemondMonnier10,Turner+14,Haworth+16}. Furthermore,
the dynamical evolution of dust grains of moderate size must be solved
in addition to the gas dynamics \citep[see reviews
by][]{Testi+14,Johansen+14}. Disks are also not necessarily in a 
steady state, and can be subject to a range of instabilities, such as
gravitational fragmentation 
\citep{Durisen+07,YoungClarke15,Forgan+15,Meru15}, the
streaming instability \citep{YoudinGoodman05}, 
Rossby wave instability
\citep[e.g.][]{Lovelace+99,Tagger01,Lyra+08b,Lyra+09a}, convective
and vertical shear instabilities, which can form and grow vortex
structures \citep[see review by][]{LyraUmurhan18}, 
the magnetorotational instability
\citep[MRI,][]{BalbusHawley91,BalbusHawley98}, as well as instabilities 
driven by non-ideal MHD \citep{Kunz08,Lesur+14}. %A summary of some of the key processes that theoriticians attempt to capture in disks is given in Fig.~\ref{fig:scheme}. 

Given the importance of these links, ultimately one wishes to identify
which physical processes affect each other in a non-negligible
fashion, and to model all of them simultaneously. The modelling of
protoplanetary disks is therefore a daunting task. Each physical
mechanism requires sufficient rigor and detail that modelling them
constitutes an active field of protoplanetary disk research in its own
right \citep[for reviews of physical processes in protoplanetary
disks, see
e.g.][]{Hartmann+98,Armitage11,WilliamsCieza11,Armitage15}. In
practice, we have neither the numerical tools nor computational
resources to achieve complete multi-physics modelling of protoplanetary disks in the immediate future.
However, we can set out a roadmap towards this goal whilst outlining achievable milestones along the way.   

\end{justify}

\subsection{Unsolved questions in computational planet formation} 

\subsubsection{The nature of accretion} 
\label{sect:accretion}

\begin{justify}

Despite major efforts, we still can not answer a central question about accretion disks: how do they accrete? Since the landmark 
work of \citet{ShakuraSunyaev73} the mechanism has been thought to be
turbulence, with the MRI identified in the early 1990s as a plausible
candidate to drive it. 
For disks hot enough to be fully ionized the MRI seems to be the most promising mechanism. However, even these disks 
 have not conclusively been shown to be appropriately magnetized.
The situation is even more dire in protoplanetary disks, as these disks are cold and
   poorly ionized, so 
the MRI is not a plausible mechanism, except for the hot inner disk ($\leq$0.1 AU), and perhaps the low density atmosphere in the outer disk, if ionization by cosmic rays and stellar ultraviolet and X-rays is sufficiently high. The rest of the disk is poorly ionized, and thus non-ideal MHD has to be taken into account. This divides the disk into a 
   small region in which MRI is active, and a large dead
zone, itself split into regions 
dominated by Ohmic, Hall, and ambipolar diffusion. The current decade has seen the realization that 
angular momentum  actually may be primarily transferred
through magnetocentrifugal winds launched from upper atmosphere
regions dominated by ambipolar diffusion
\citep{BaiStone13,Bai13,Gressel+15,Simon+18}. Some disks are seen to have
free-free emission \citep[GM Aur,][]{Ricci+10,Owen+13,Macias+16} but it
is not clear whether the wind is photoevaporative 
\citep{Owen+13,Canovas+18} or magnetocentrifugal
\citep{Banzatti+19}. The Hall-dominated region is prone to a
Hall-shear instability \citep{Kunz08}, that would drive laminar
accretion \cite{Lesur+14}. Finally, the Ohmic region has three
possible instabilities: the Zombie Vortex Instability
\citep{Marcus+15,Barranco+18}, the Vertical Shear Instability
\citep{Nelson+13,Flock+17}, and the Convective Overstability
\citep{KlahrHubbard14,Lyra14}, all generating large scale vortices and
moderate levels of turbulence. 

Observational tests to determine whether disks are magnetized, the prevalence of winds, and
the level of turbulence throughout the whole disk column should provide the necessary information to decide between these models. Computational models with the necessary physics and resolution to solve for all these dynamical instabilities together is sorely missing but possible sometime early in the next decade. Such models will inform the observations of the main dynamical processes expected in disks and where to expect them.   

\end{justify}

\subsection{Ab initio planet formation}  

\begin{justify}

Planetesimals and planetary embryos are $\leq 10^3$~km in size, 
   suggesting that direct observations of their formation will remain difficult in the next decade.
On this front, the best data comes from Solar System missions to 
    primitive 
asteroids and comets such as provided by Rosetta, New Horizons,
   Hayabusa 2, OSIRIS-REx, and Lucy.
Planetesimal formation by streaming instability
\citep[SI,][]{YoudinGoodman05,Johansen+07}, followed by pebble
accretion to planetary embryos
\citep{OrmelKlahr10,LambrechtsJohansen12} stands as the best candidate
for the formation of rocky and icy planets. Pebble trapping in
vortices \citep{BargeSommeria95,Lyra+08b,Lyra+09a} is also a plausible
candidate, and can be tested observationally, as the spatial scales
are resolved in some disks (Oph IRS 48, \citealt{vanderMarel+13}, and MWC 758, \citealt{Casassus+18}). These hypotheses should be seen as complementary rather than mutually exclusive. Models still have problems making the SI work with solar metallicity \citep{Carrera+15}, needing at least twice the amount of metals (a small but numerically robust factor). In real disks, it is likely the streaming instability feeds off dust concentration in
local pressure maxima such as rings or vortices \cite[e.g.][]{Lyra+08b,Raettig+15}. 
A detailed global model of streaming instability should be able to address these questions. What is needed is a model with sufficient dynamical range (1--100 AU), and enough resolution (at least 20 grid zones per SI wavelength), with full 360$^\circ$ azimuthal coverage in the MRI-dead zone in the presence of hydrodynamical instabilities. Other questions pertain to 
the angular momentum distribution of clumps formed by SI
    during gravitational collapse, as initially investigated by \citet{Nesvorny+10}. 
In particular, the bizarre shape of Ultima Thule, as well as the frequency of contact binaries 
   among small Solar System bodies
are problems that have 
   been emphasized
to the community after the Rosetta and New Horizons encounters, and will take effort on the computational side in the next decade to explain.   

\end{justify}

\subsection{Disk formation and early evolution}  

\begin{justify}

There is growing evidence for planet formation happening early. How
then, are disks assembled and what progress towards dust
processing/planet formation is made at this early stage? How
does the star forming environment sculpt the disk parameters and can
this imprint upon planetary populations? Unfortunately, there are too
many free parameters to clearly connect the environment of molecular clouds to the
initial conditions for circumplanetary disks. Although disk formation 
has been an active area of research \cite[e.g.][]{Inutsuka12,Tsukamoto16, WursterLi18}
we lack observational data on the first 10$^4$ years of their life when evolution is fast. 

A possible relevant process in this early stage is stellar flybys 
that affect not only disk truncation but also disk warping and morphology \citep{Cuello+19}. The younger the cluster, the
more likely the encounters, which can strongly
affect where and when planetesimals form. This is also particularly important for the
gravitational instability model of giant planet formation, which would have
to happen early, while the disks are massive. While
probably not the dominant mode of planet formation, some planetary
systems are indeed extremely difficult to explain by core accretion.

\end{justify}

\subsubsection{Circumplanetary disks}  

\begin{justify}

As disk observations advance in accuracy and detail, the goal of
imaging a planet in the making has been achieved, around the star PDS70 \citep{Keppler+18}. 
While proving a unequaled laboratory to study planet-disk interaction, it remains 
unclear whether the emission originates from the planet’s photosphere
or from a circumplanetary disk (CPD) surrounding the protoplanet. CPDs
are formed in high resolution numerical simulations around giant
planets, and are a natural explanation for the regular satellite
system of Jupiter. A gas CPD of about half the Hill radius should
exist as long as the circumstellar disk (CSD) exists, because the CSD
continuously feeds the CPD \citep{Szulagyi+14, FungChiang16}. Modeling observations of CPDs is challenging, requiring 3D
radiation-hydrodynamics with adaptive mesh refinement. The
post-processing is also challenging, since the temperatures and
opacities remain poorly known. Another possible way to detect them is
via kinematics: signposts of circumplanetary kinematics have been predicted from 3D
hydrodynamic simulations of CPDs as localized deviations from
Keplerian velocity in protoplanetary disks \citep{Perez+15}, 
and have been later detected in ALMA observations of the HD163296 disk
\citep{Pinte+18}. Modeling of CPDs is an area that we recommend 
significant effort in the next decade. 

\end{justify}

\section{Recommendations}

\subsection{Approximations}
\begin{justify}

Long a popular closure model for turbulence, the $\alpha$ viscosity model \citep{ShakuraSunyaev73}
has outgrown its usefulness. Its use in 2D and 3D modeling has become
a hindrance to true progress in the field. Disks are not viscous, they are inviscid.
It has been established that $\alpha$ viscosity, as a mean-field 
theory for disc turbulence, does not capture the spatial distribution and spectral properties now understood, particularly those new mechanisms of 
Sect.~\ref{sect:accretion}. As such, models with $\alpha$ viscosity result in incorrect small scale properties.
This has a major
impact on structure formation, the location of planetesimal formation, the shape of planet-induced gaps and vortices.
Even the planet-disc interactions and planet migration torques can be impacted. Thus, a major goal for the next decade is to understand the small-scale structure well enough to develop subgrid models suitable for use in large eddy simulations. This will allow the large-scale flows to be modeled without extraordinary 
computational resources.

Another point (a dismal one to make as late as 2019), is 
   the need to move beyond
isothermal models. Already in the decade of 2000--2009 major
advances were made in planet formation by relaxing this crippling
approximation \cite[e.g., outward migration,][]{PaardekooperMellema06}. The same happened in 2010--2019. For the observations,
especially because the disk scale height controls the scattered
intensity in infrared/optical, it is crucial that models that intend
to reproduce observations do so both in 3D and 
    without using the isothermal approximation
\citep{FungDong15,Hord+17}. There is little value in a model,
even if it does a reasonable job at reproducing observations, if it is
intrinsically flawed. The adoption of $\alpha$ viscosity and isothermal
thermodynamics are approximations
that we can and should get away from during the next decade.

\end{justify}

\subsection{Evolutionary models and population synthesis} 

\begin{justify}

Evolutionary models must
run for the entire lifetime of the disk ($\sim$10 Myr). Such models are currently used for
population synthesis, 
with the goal of comparing results to the observed distribution of exoplanets
\citep[e.g.][]{Mordasini+09a,Mordasini+09b,IdaLin10,Dittkrist+14,Mordasini18,Ndugu+18}. These models 
have also used N-body for the interacting planets
\citep{Horn+12,ColemanNelson16,Lambrechts+19,Izidoro+19,Bitsch+19}, 
but with decoupled 1D hydrodynamics for the disk, 
using formulae for planet-disk interaction derived from hydrodynamical simulations 
with more degrees of freedom \citep{Paardekooper+11}. 
The pitfalls of such an approach are many and dangerous. 

We identify these models of planetary system evolution as a field 
would benefit from a significant overhaul. A 2D hydrodynamic model of a disc and planets over 10~Myr timescales is already
possible with current computational resources if the inner 
boundary is at a few AU. 
However, this has been done with either massively parallel use of CPUs, 
or only moderately parallel graphical processing units (GPUs)
\citep{Regaly+12,Muley+19}. \cite{Perez+18} report 20 orbits/hr
in 3D with 4 GPUs using direct GPU-GPU communication. Extrapolating the speedup in GPU-based computation of a order of 
magnitude in the last decade is continued into the next, 
a new generation of evolutionary models directly modelling 
disc-planet interactions over the full lifetime and 
spatial scales of the disk may be built.

These models can then be used to make significant advancements such as
self-consistently including photoevaporation clearing of the disc gas,
or used for the basis of a new generation of 
high-fidelity population synthesis models.
In particular, this second goal, requiring coverage of a 
vast parameter space, may be enabled by the increasing speed and 
decreasing cost of GPU-type computing devices in the future.

\end{justify}

\section{Conclusion}  

\begin{justify}

We have identified major fields of interest in computational planet formation: the nature of accretion, ab initio planet formation, early evolution, and circumplanetary disks. 
      We recommend that modelers relax the approximations of $\alpha$ viscosity and isothermal equations of state,
on the grounds that 
these models use flawed assumptions, even if they give good visual qualitative agreement with observations. We similarly recommend that population 
synthesis move away from 1D hydrodynamics. The computational 
     resources to reach these goals should be developed during the next decade, through improvements in algorithms and the hardware for hybrid CPU/GPU clusters.
Advances in computational planet formation, coupled with high angular resolution and great line sensitivity in ground based 
   interferometers, 
ELTs and JWST, should allow for large strides in the field in the next decade.

\end{justify}

\pagebreak
%\textbf{References}

%\bibliography{master.bib}
%\bibliography{master.bib}
\bibliography{main.bbl}

\end{document}